\documentclass[11pt]{article}
\usepackage{vietnam,epsfig}

\def\Journal#1#2#3#4{{#1} {\bf #2}, #3 (#4)}

\def\PRL{\em Phys. Rev. Lett.}

\def\ipb{\ensuremath{\mathrm {pb}^{-1}}}
\def\ifb{\ensuremath{\mathrm {fb}^{-1}}}
\def\W{\mbox{$\mathrm{W}$}}

\def\MW{\mbox{$m_{\mathrm{W}}$}}

\def\GEV{\mbox{$\mathrm{GeV}$}}

\def\GEVcc{\mbox{$\mathrm{GeV}/{{\it c}^2}$}}
\def\MEVcc{\mbox{$\mathrm{MeV}/{{\it c}^2}$}}
\def\epem{\mbox{$\mathrm{e}^+\mathrm{e}^-$}}
\def\rs{\mbox{$\sqrt{s}$}} 
\def\qq{\mbox{$\mathrm{qq}$}}

\def\ev{\mbox{$\mathrm{e}\nu$}}
\def\mv{\mbox{$\mu\nu$}}
\def\tv{\mbox{$\tau\nu$}}
\def\lv{\mbox{$\ell\nu$}}

\def\ra{\rightarrow}

\def\be{\begin{equation}}
\def\ee{\end{equation}}
\def\bea{\begin{eqnarray}}
\def\eea{\end{eqnarray}}

\begin{document}
\vspace*{4cm}
\title{W PHYSICS AT LEP2}

\author{P. AZZURRI}

\address{Scuola Normale Superiore, 
Piazza dei Cavalieri 7, 56100 Pisa, Italy}

\maketitle\abstracts{
In five years of operation (1996-2000) the four 
LEP2 experiments 
collected roughly 3~\ifb of data from $\epem$ collisions 
at centre-of-mass energies between 161 and 207~$\GEV$,
yielding a total of about 40000 W-pair events.
The analysis of the LEP2 W-pair production rates and kinematics 
has allowed fundamental new measurements of the electroweak model 
to be performed, most notably: (i) the first determination of 
gauge boson self couplings, (ii) direct measurements of W
decay couplings to all fermions, and (iii) a direct measurement of the W mass
with a 40~$\MEVcc$ precision, 
comparable to hadron collider and indirect determinations. 
}

\section{W-pair cross section}
\begin{figure}[htb]
  \centerline{\epsfig{file=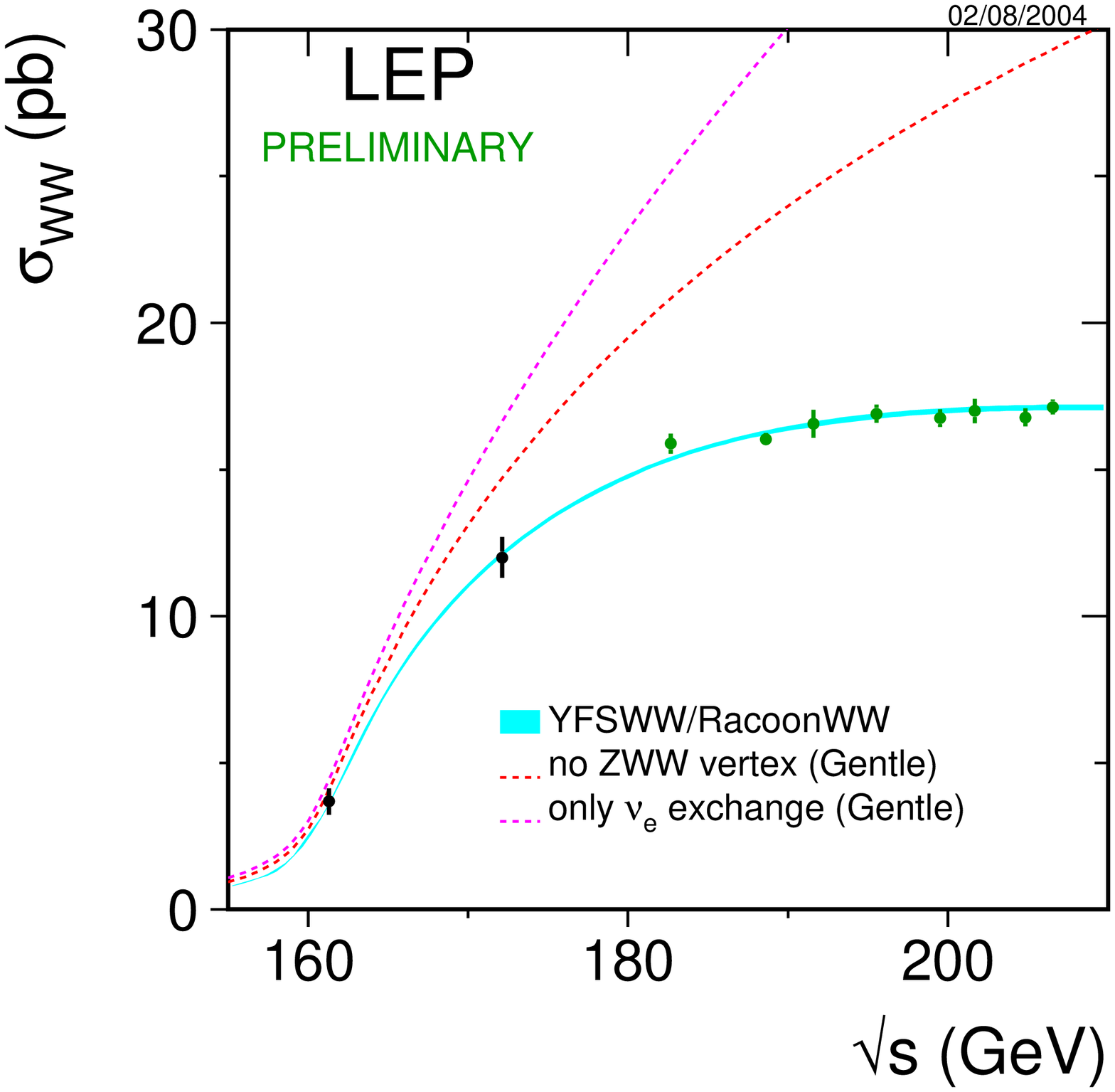,height=9cm}
              \epsfig{file=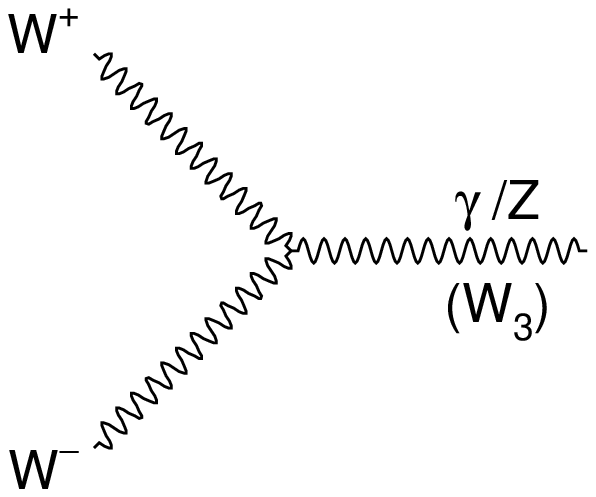,height=6cm}}
\caption{\sl Total W-pair cross-sections measured at LEP 
at $\rs$=161-207~GeV, compared to SM predictions. 
The two dashed curves show the predicted W-pair 
cross section in the absence of the WWZ and WW$\gamma$ couplings
shown on the right.}
\label{fig:xsec}
\end{figure}

Pair-production of W bosons is of great importance 
in the Standard Model (SM) 
of electroweak interactions~\cite{SM}, where the 
non-abelian nature of the SU(2) group leads to 
three-line (TGC) boson vertices that play a crucial role
in the W-pair production.
The LEP2 data sample of ~700$~\ipb$ per experiment at $\rs$=161-207~GeV
allowed to collect about 10$^4$ W-pairs per experiment,
identifying these events in all their final states, 
leptonic and hadronic. 
The results for the total W-pair 
cross sections as a function of the 
centre-of-mass energy~\cite{ew}
are shown in Fig.~\ref{fig:xsec},
and are in overall 
agreement with SM predictions at the level of 1\%.
As it can be seen, the LEP2 W-pair cross 
section measurement alone represents a stunning proof of the 
presence of both the WWZ and WW$\gamma$ couplings dictated by the
electroweak SU(2)$\otimes$U(1) gauge structure.

\section{W couplings to gauge bosons}
The structure and magnitude of WWZ and WW$\gamma$ couplings
are measured fitting the W-pair event rates and the angular 
distributions of the W production and decay axes~\cite{ew}.
The most general Lorentz-invariant WWV vertex (V=$\gamma$,Z)
can have seven complex couplings
$$
g_1^{\rm V} \;\; \kappa_{\rm V} \;\; \lambda_{\rm V} \;\; 
g_5^{\rm V} \;\; g_4^{\rm V} \;\; 
\tilde{\kappa}_{\rm V} \;\; \tilde{\lambda}_{\rm V}
$$
making up a total of 28 parameters for both WWZ and WW$\gamma$
vertices. In the SU(2)$\otimes$U(1) model we expect 
$g_1^{\rm V}=\kappa_{\rm V}=1$ for both V=$\gamma$ and V=Z,
while all other real and imaginary parts are expected to vanish.
A fit to ALEPH data only 
to all 28 parameters~\cite{atgc} leads to 
\begin{eqnarray*}
 {\rm Re} (g_1^{\gamma})= 1.123\pm 0.091   &  &
  {\rm Re} (\kappa_{\gamma})= 1.071 \pm 0.062 \\
 {\rm Re} (g_1^{\rm Z})= 1.066 \pm 0.073  &  &
  {\rm Re} (\kappa_{\rm Z})= 1.065 \pm 0.061   
\end{eqnarray*}
and all other 24 coupling parameters consistent with zero, within
uncertainties ranging from 0.035 to 0.250. 
The result of this fit shows how clearly the W-pair data 
has revealed the SU(2)$\otimes$U(1) structure of
the gauge self-couplings.

A more constrained fit of all LEP2 data,
in search of anomalous contributions to TGC,
 to the three couplings that conserve 
separately C and P, U(1)$_{\rm em}$, and global SU(2)$_L\otimes$U(1)$_Y$,
yields~\cite{ew}
\begin{eqnarray*}
  \kappa_{\gamma}= 0.984 \pm 0.045 \; &
  \lambda_{\gamma}= -0.016 \pm 0.022 \;&
  g_1^{\rm Z}= 0.991 \pm 0.021 ,  
\end{eqnarray*}
revealing again no deviation from the SM expectations.

\section{W decay couplings}
Given the possibility to classifying all W decays modes, 
the LEP2 W-pair sample has allowed
the first direct measurement of all leptonic and hadronic
W decay branching ratios, as 
\begin{eqnarray*}
{\rm B}(\W\ra\qq )= 67.49 \pm 0.28 \% \\
{\rm B}(\W\ra\ev)= 10.66 \pm 0.18 \%  \\
{\rm B}(\W\ra\mv)= 10.60 \pm 0.15 \%  \\
{\rm B}(\W\ra\tv)= 11.41 \pm 0.22 \%
\end{eqnarray*}
where the hadronic decay fraction {\rm B}(W$\ra\qq$)
is determined under the assumption of 
lepton coupling universality 
${\rm B}(\W\ra\ev)= {\rm B}(\W\ra\mv)={\rm B}(\W\ra\tv)= 
\left( 1- {\rm B}(\W\ra\qq )\right)/3$.
In the case of lepton non-universality,
it can be noticed that the tau decay fraction is currently
about three sigmas larger than the electron-muon 
average.

The above results can be interpreted as 
a test of the lepton-quark universality 
of charged currents 
($g_{\rm q}/g_\ell= 1.000\pm 0.006$), 
and of the lepton family universality 
of charged currents 
($ g_\mu/g_{\rm e} = 0.997\pm 0.010$,  
 $ g_\tau/g_{\rm e} =  1.034\pm 0.015$,  
 $ g_\tau/g_{\mu} =  1.037\pm 0.014 $).
The W hadronic decay fraction can also be interpreted as a test
of the unitarity of the CKM quark mixing matrix, in the 
first two families as 
$\sum |V_{ij}| (i=u,c ;\; j=d,s,b)= 2.000\pm 0.026$, and from 
this extract the W$cs$ coupling amplitude 
$|V_{cs}|=0.976\pm 0.014$,
without CKM unitarity assumptions. 

\section{W mass}
\subsection{W mass from threshold cross section}
At the start of LEP2 about 10~$\ipb$ of data per
experiment were recorded near the W-pair production threshold 
($\rs\simeq 161~\GEV$), where the production cross section
alone is very sensitive to the $\MW$ value. From the 
cross section determination an independent measurement 
of the W mass has been obtained to be
$ \MW=80.40\pm 0.20 ~\GEVcc $, 
where the large uncertainty is due to the limited statistics 
of the data sample.
\subsection{W mass from kinematic reconstruction}
\begin{figure}[htb]
  \centerline{\epsfig{file=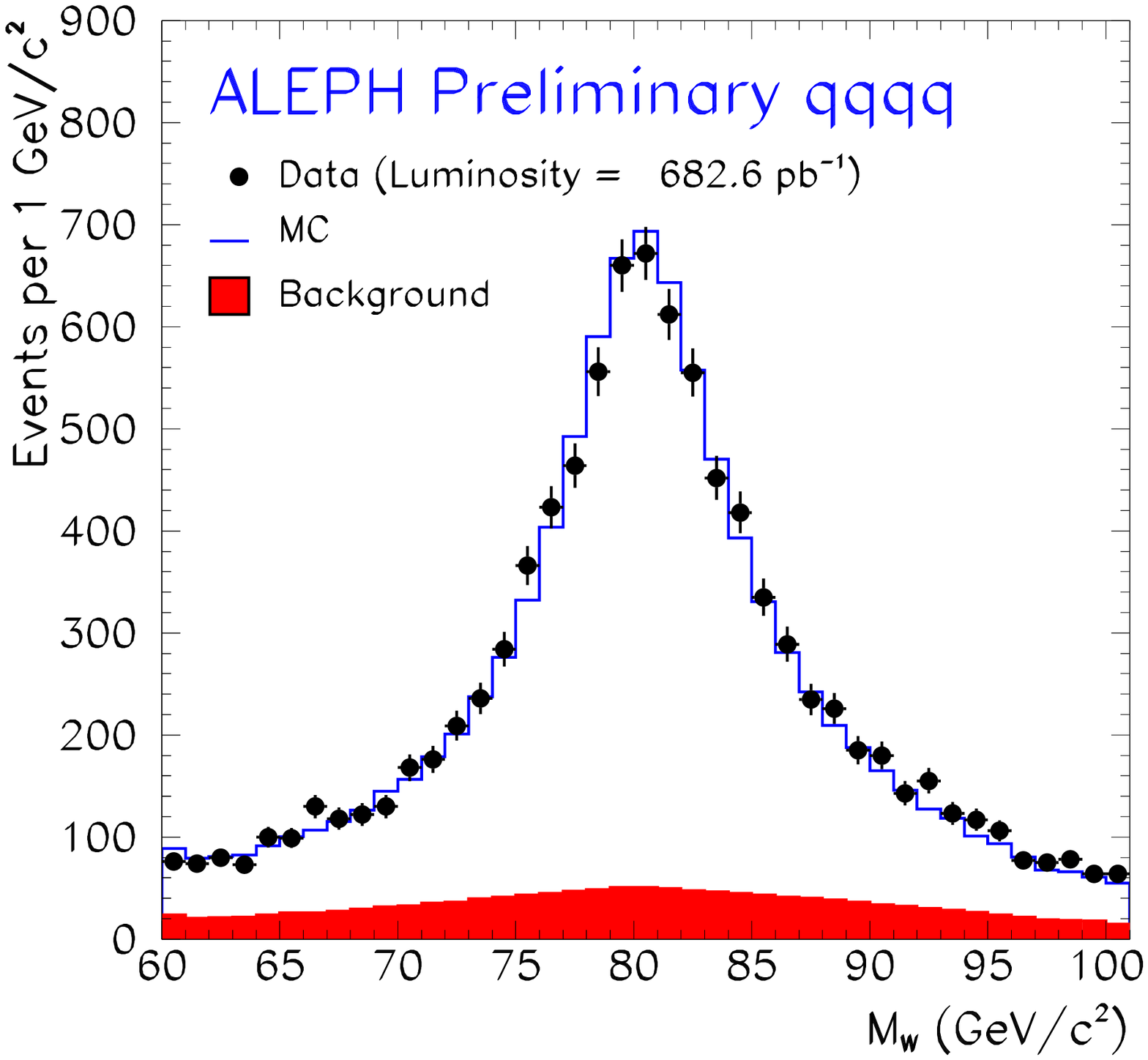,height=8cm}
              \epsfig{file=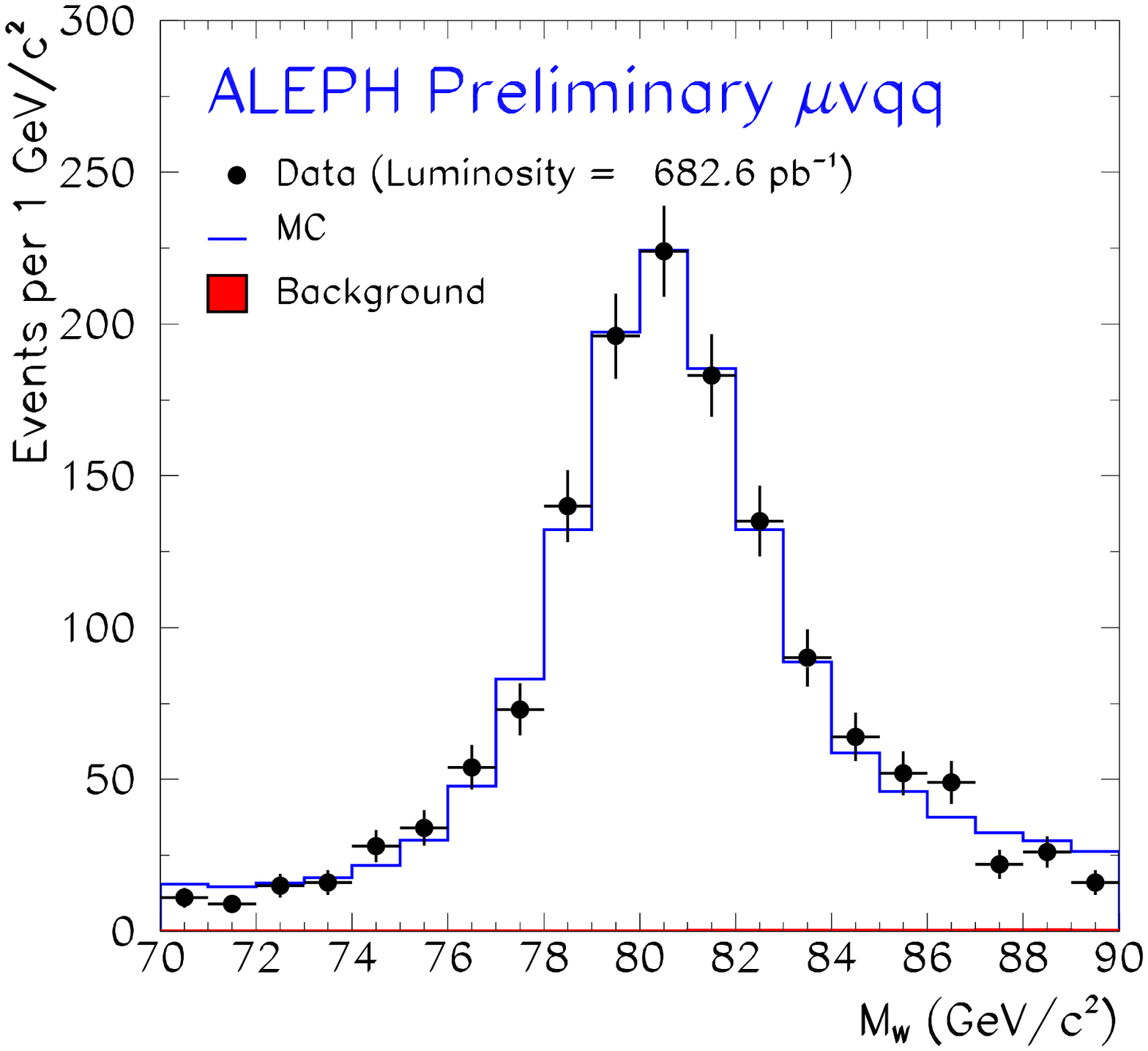,height=8cm}}
\caption{\sl ALEPH reconstructed W mass distributions after applying
energy-momentum conservation constraints, in 
the fully hadronic channel (left), and in the semi-leptonic
muon channel (right).}
\label{fig:mw}
\end{figure}

The W invariant mass is reconstructed event-by-event
in all $\qq\qq$ and $\qq\lv$ decays of W-pairs,
from the kinematics of the visible decay particles.
The resolution of the W mass peak is improved 
by applying a kinematic fit imposing 
energy-momentum conservation constraints, 
leading to mass distributions as those shown in Fig.~\ref{fig:mw}.

\begin{figure}[htb]
  \centerline{\epsfig{file=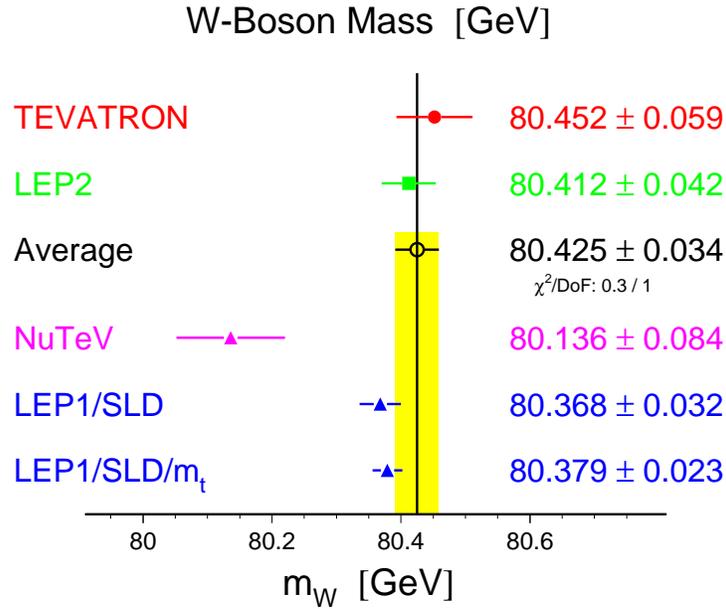,height=8cm}}
\caption{\sl Comparison of W mass measurements. Results from the 
direct measurements from TEVATRON and LEP2 
data~\protect\cite{ew} are shown on the top.
Indirect constraints from other electroweak measurements are shown 
on the bottom~\protect\cite{ew}. 
Separately, the NuTeV determination~\protect\cite{nut} 
is 2.8 sigmas lower than the direct determinations.}
\label{fig:av}
\end{figure}

The $\MW$ value is extracted from the W mass distributions
with different methods as fits with a 
probability density function (p.d.f.) calibrated with Monte~Carlo
simulations, or Monte~Carlo re-weighting techniques using the 
measured masses and their errors as inputs~\cite{ew}.
The $\MW$ values and systematic uncertainties are 
evaluated separately for each W-pair decay topology,
($\qq\qq$, $\ev\qq$, $\mv\qq$ and $\tv\qq$) and the final 
$\MW$ values is obtained by combining the measurement from the 
individual channels. The most recent combined result from the LEP2 data is
$$
\MW=80.412\pm 0.042 ~\GEVcc 
$$
where the weight of the fully hadronic ($\qq\qq$) channel
is only 10\% because of large uncertainties coming from 
possible final state interactions (FSI) effects between 
the two W hadronic decay products.
Recent prospects to reduce the FSI uncertainty on $\MW$
in the hadronic channel from 100 to 40-50~$\MEVcc$
should bring down the combined error on $\MW$ from 42 to 35-38~\MEVcc.
The current LEP2 determination and other direct and indirect 
measurements of $\MW$ are shown in Fig.~\ref{fig:av}.

\end{document}
